\documentclass[aps,prd,preprint,groupedaddress,showpacs]{revtex4}
\usepackage[dvipdfm]{graphicx}
\usepackage{amsmath}

\newcommand{\Slash}[1]{\ooalign{\hfil/\hfil\crcr$#1$}}
\begin{document}
\title{Pionic BEC--BCS crossover at finite isospin chemical potential}

\author{Masayuki Matsuzaki}
\email[]{matsuza@fukuoka-edu.ac.jp}
\affiliation{Department of Physics, Fukuoka University of Education, 
             Munakata, Fukuoka 811-4192, Japan}

\date{\today}

\begin{abstract}
 We study the character change of the pionic condensation at finite isospin 
chemical potential $\mu_\mathrm{I}$ by adopting the linear sigma model as a 
non-local interaction between quarks. At low $|\mu_\mathrm{I}|$ the condensation 
is purely bosonic, then the Cooper pairing around the Fermi surface grows 
gradually as $|\mu_\mathrm{I}|$ increases. This $q$-$\bar q$ pairing is weakly 
coupled in comparison with the case of the $q$-$q$ pairing that leads to 
color superconductivity. 
\end{abstract}

\maketitle
 Recent progress in computer power makes it possible to reliably simulate 
quantum chromodynamics (QCD) at finite temperature $T$. As for finite density 
(usually parametrized by finite baryon chemical potential $\mu_\mathrm{B}$), 
however, the well known sign problem limits simulations. Alternatively, 
QCD at finite isospin chemical potential $\mu_\mathrm{I}=\mu_u-\mu_d$ (where 
$\mu_u$ and $\mu_d$ denoting the chemical potential of $u$ and $d$ quark, 
respectively) as well as the SU(2) color systems, in which the sign problem 
does not exist, are studied to give insights into the actual finite 
$\mu_\mathrm{B}$ physics~\cite{KS}. These systems are also studied extensively 
in terms of effective models~\cite{FBO,Ni,BCPR,LV,MMR,ZL,AB}. One of the most interesting 
aspects of the finite $\mu_\mathrm{I}$ systems is that they accommodate pion 
condensation for $|\mu_\mathrm{I}|>m_\pi$~\cite{CDM}, with $m_\pi$ denoting 
the mass of pions. Son and Stephanov~\cite{SS} predicted that the pion condensed 
phase evolves to Cooper pairing between $u$ and $\bar d$ ($d$ and $\bar u$) 
for $\mu_\mathrm{I}>0$ ($<0$) at high $|\mu_\mathrm{I}|$, but the quantitative 
process of the character change of the condensation has not been discussed. 

 The BEC--BCS crossover has long been expected to occur in various quantum 
systems~\cite{Le,NS,HW}; it was experimentally observed in ultra cold atomic gases, 
in which the strength of the interaction can be tuned artificially, only 
recently. 
At least in principle, it can occur also in systems governed by the strong 
interaction, in which the strength of the interaction can not be tuned artificially 
aside from theoretical simulations~\cite{NA}. Rather, the change in the 
environment, typically density, would lead to the crossover~\cite{SHZ}. 
In symmetric nuclear matter, the neutron ($n$)--proton ($p$) pairing in the 
$^3S_1$ + $^3D_1$ channel that leads to bound deuteron formation was studied in 
Ref.~\cite{BLS}. The $n$--$n$ and $p$--$p$ $^1S_0$ pairing, that has attracted 
attention from viewpoints of both nuclear structure and neutron stars, however, 
does not reach the BEC~\cite{Tani,Matsuo}. In intermediate density quark 
matter, the present author discussed that the spatial extension of quark Cooper 
pairs in a color superconductor is comparable with the mean interparticle 
distance~\cite{MM_q}. Later, a wide enough density region was studied~\cite{AHI} 
and it was shown that the diquark pairing becomes weak at extremely high density. 
The properties of the pseudo gap phase and bosonic excitations were 
studied in Refs.~\cite{KKKN,CNZ,Br}. 

 Since the mechanism of the fermion-antifermion condensation that produces the 
fermion mass is essentially the same as the BCS pairing as recognized in 
Nambu and Jona-Lasinio's celebrated paper~\cite{NJL}, the evolution of the 
charged pion condensation to $q$--$\bar q$ Cooper pairs can be analyzed in the 
context of the BEC--BCS crossover in terms of the spatial structure of the pion 
condensation. To this end, one must introduce a non-local interaction between 
$q$ and $\bar q$ that gives momentum dependent condensations. In the present 
study, we adopt the linear sigma model~\cite{GL}, which respects chiral symmetry, 
as an inter-quark interaction, since 1) the pion condensation occurs as a 
spontaneous symmetry breaking among three pions that have light but non-zero 
masses after the chiral symmetry breaking between the sigma meson and the pions, 
and 2) the effect of high $|\mu_\mathrm{I}|$ on it has long been 
studied~\cite{CDM,HJZ,MPSZ,An}. In Ref.~\cite{DSW} the BEC--BCS crossover in the 
diquark pairing was studied in a boson--fermion model similar to that of the present 
study but the condensation is momentum independent. 

 Finite $\mu_\mathrm{I}$ occurs with finite $\mu_\mathrm{B}$ in the real world; 
with finite $T$ and small $\mu_\mathrm{B}$, for example 0.04 GeV~\cite{BF}, 
in heavy ion collisions and with (near) zero $T$ and large $\mu_\mathrm{B}$, 
for example $\agt$ 1 GeV, in compact stars. 
In this sense, the present study of the system with $\mu_\mathrm{B}=0$ is 
just the first step to investigate the realistic systems. 
However, since a signature of the BEC--BCS crossover in the chemical potential 
dependence of the condensation is measured in a lattice simulation for the SU(2) 
color system~\cite{HKS} that is in a sense dual~\cite{KS2} to the finite 
$\mu_\mathrm{I}$ system, the spatial structure 
of the composite pions would be worth studying even with $\mu_\mathrm{B}=0$. 

 When a conserved charge density $\mathcal{N}$ exists, the effective Lagrangian 
density is obtained with replacing the Hamiltonian density $\mathcal{H}$ by 
$\mathcal{H}-\mu\mathcal{N}$, here $\mu$ denoting the corresponding chemical 
potential, in the partition function and performing momentum-field 
integrations~\cite{Ka}. The result for the charged pion is 
\begin{equation}
\mathcal{L}_\mathrm{eff}=\mathcal{L}
(\dot{\pi_1}\rightarrow\dot{\pi_1}-\mu\pi_2,
 \dot{\pi_2}\rightarrow\dot{\pi_2}+\mu\pi_1) .
\end{equation}
Since the isospin chemical potential $\mu_\mathrm{I}$ corresponds to the charge 
chemical potential in the hadronic world, this form applies to the present purpose. 
This indicates that the role of $\mu_\mathrm{I}$ corresponds to that of the angular 
frequency in the non-relativistic spatial rotation, that is, to move to a 
``coordinate frame" rotating in the 3 dimensional isospin space; the zero-energy 
rotational motion is a physical image of the Nambu--Goldstone mode. 

 The adopted effective Lagrangian for the quarks, sigma mesons and pions is 
\begin{gather}
\mathcal{L}_\mathrm{eff}=\mathcal{L}_\mathrm{q}+\mathcal{L}_\mathrm{M}+\mathcal{L}_\mathrm{couple} ,
\notag \\
\mathcal{L}_\mathrm{q}=\bar q(i\Slash{\partial}-m_q+\frac{\mu_\mathrm{I}}{2}\gamma^0\tau_3)q ,
\notag \\
\mathcal{L}_\mathrm{M}
=\frac{1}{2}(\partial_\mu\sigma\partial^\mu\sigma
            +\partial_\mu\overrightarrow{\pi}\cdot\partial^\mu\overrightarrow{\pi})
            -U(\sigma,\overrightarrow{\pi})
\notag \\
+\mu_\mathrm{I}(\pi_1\dot{\pi_2}-\pi_2\dot{\pi_1})+\frac{\mu_\mathrm{I}^2}{2}(\pi_1^2+\pi_2^2) ,
\notag \\
U(\sigma,\overrightarrow{\pi})
=\frac{\lambda^2}{4}(\sigma^2+\overrightarrow{\pi}^2)^2
-\frac{m_0^2}{2}(\sigma^2+\overrightarrow{\pi}^2)
-c\sigma ,
\notag \\
m_0^2=\lambda^2f_\pi^2-m_\pi^2\quad(>0), \quad
c=f_\pi m_\pi^2 ,
\notag \\
\mathcal{L}_\mathrm{couple}
=-G\bar q(\sigma+i\gamma^5\overrightarrow{\tau}\cdot\overrightarrow{\pi})q ,
\end{gather}
where $f_\pi$ and $m_\pi$ stand for the pion decay constant and the pion mass, 
respectively. Hereafter, quantum fluctuations are indicated by primes, such as, 
\begin{gather}
\bar q\gamma^\mu q=\langle\bar q\gamma^\mu q\rangle+(\bar q\gamma^\mu q)' , 
\notag \\
\sigma=\langle\sigma\rangle+\sigma' , \quad 
\pi_i=\langle\pi_i\rangle+\pi_i' .
\end{gather}
Since the quantum fluctuations of the quark densities and the meson fields after 
subtracting the mean field couple to each other, the normal product in 
$\mathcal{L}_\mathrm{eff}$ is understood. 
Note here that charge neutrality forced by electrons are often considered in 
studies of realistic $\mu_\mathrm{B}\neq0$ matter expected to exist in compact 
stars~\cite{EK,Abu}. In the present study, 
however, charge neutrality is not forced since the asymmetric ($\mu_\mathrm{I}\neq0$) but 
$\mu_\mathrm{B}=0$ system is an idealized one from the beginning. 
On the other hand, the charge introduced by 
$\mu_\mathrm{I}$ is conserved among quarks and mesons. 

 It is well known that, in the mean field level, 
$U_\mathrm{eff}=U(\sigma,\overrightarrow{\pi})-\frac{\mu_\mathrm{I}^2}{2}(\pi_1^2+\pi_2^2)$ 
has the minimum at 
\begin{equation}
\langle\sigma\rangle=\frac{f_\pi m_\pi^2}{\mu_\mathrm{I}^2},\quad 
\langle\pi\rangle^2=\frac{\mu_\mathrm{I}^2-m_\pi^2}{\lambda^2}+f_\pi^2-\langle\sigma\rangle^2
\end{equation}
for $|\mu_\mathrm{I}|>m_\pi$, assuming $\langle\pi_3\rangle=0$~\cite{CDM,HJZ}. 
We take $\langle\pi_1\rangle=\langle\pi\rangle$ and $\langle\pi_2\rangle=0$ without 
loss of generality. 
This means that the pion condensation exists in both charge sectors 
irrespective of the sign of $\mu_\mathrm{I}$. 
After expanding $\mathcal{L}_\mathrm{M}$ up to the quadratic 
terms in $\sigma'$ and $\pi_i'$, diagonalization of the coupled Klein-Gordon equations 
for $\sigma'$, $\pi_1'$ and $\pi_2'$ gives the mass eigenvalues, one of which is zero 
as done in Ref.~\cite{HJZ}. But the meson mixing can not be calculated since the 
$3\times3$ mass matrix is not regular. Thus, another approximation must be sought. 
Note that the meson mixing was calculated in another model~\cite{HZ}. Since the essential 
character of the massless meson propagation in the pion condensed phase is 
the rotational motion 
in the isospin space, we adopt a polar coordinate representation, 
\begin{equation}
\pi_\pm=\frac{1}{\sqrt{2}}(\pi_1\pm i\pi_2)=\frac{1}{\sqrt{2}}\pi\exp{(\pm i\theta)}
=\frac{1}{\sqrt{2}}(\langle\pi\rangle+\pi')\exp{(\pm i\theta)} ,
\end{equation}
without expanding the angular field. This representation assures the conservation of 
the (third component of the isospin) current of the total system seen in the ``rotating" 
frame: 
\begin{equation}
\partial_\mu j^\mu=\partial_\mu(\bar q\gamma^\mu\frac{\tau_3}{2}q)
                  +\partial_\mu(\pi_1\partial^\mu\pi_2-\pi_2\partial^\mu\pi_1)
                  +\mu_\mathrm{I}\partial_t(\pi_1^2+\pi_2^2)=0 ,
\end{equation}
within the quadratic terms of the fluctuating quantum fields. In other words, the 
equation of motion of the angular field assures the current conservation. 

 After confirming this point, we write down the coupled Klein-Gordon equations retaining 
the lowest order terms in each equation as 
\begin{gather}
\partial_\mu\partial^\mu\sigma'+(2\lambda^2\langle\sigma\rangle^2+\mu_\mathrm{I}^2)\sigma'
+2\lambda^2\langle\sigma\rangle\langle\pi\rangle\pi'=-G(\bar qq)',
\notag \\
\partial_\mu\partial^\mu\pi'+2\lambda^2\langle\pi\rangle^2\pi'
+2\lambda^2\langle\sigma\rangle\langle\pi\rangle\sigma'-2\mu_\mathrm{I}\langle\pi\rangle\dot{\theta}
=-G(\bar qi\gamma^5\tau_1q)',
\notag \\
\langle\pi\rangle\partial_\mu\partial^\mu\theta=-G(\bar qi\gamma^5\tau_2q)',
\notag \\
\partial_\mu\partial^\mu\pi_3'+\mu_\mathrm{I}^2\pi_3'=-G(\bar qi\gamma^5\tau_3q)'.
\label{KG}
\end{gather}
Here we make one additional approximation to handle the set of equations: We ignore 
$-2\mu_\mathrm{I}\langle\pi\rangle\dot{\theta}$ in the second equation that corresponds to 
the Coriolis coupling. Its influence will be checked later. The obtained set contains 
1) the $\sigma$--$\pi$ mixing (the first and second equations), and 2) the rotational 
massless field (the third equation) due to the existence of the pion condensation 
$\langle\pi\rangle$. 

 The equation of motion of the quark propagator 
\begin{equation}
G_{\alpha\beta}^{ij}(x-x')
=-i\langle\tilde0\vert Tq_\alpha^i(x)\bar q_\beta^j(x')\vert\tilde0\rangle, 
\label{propagator}
\end{equation}
where $i$, $j$ and $\alpha$, $\beta$ represent isospin and Dirac indices, respectively, 
and $\vert\tilde0\rangle$ is the pion condensed ground state, is given by 
\begin{equation}
\begin{split}
&(i\Slash{\partial}-m_q+\frac{\mu_\mathrm{I}}{2}\gamma^0\tau_3)G(x-x')
\\
&=\delta^4(x-x')
-iG\langle\tilde0\vert T(\sigma(x)+i\gamma^5\overrightarrow{\tau}
                                  \cdot\overrightarrow{\pi}(x))
            q(x)\bar q(x')\vert\tilde0\rangle .
\label{EOM}
\end{split}
\end{equation}
After sorting the mean field terms in 
\begin{equation}
\sigma+i\gamma^5\overrightarrow{\tau}\cdot\overrightarrow{\pi}
\simeq
\langle\sigma\rangle+i\gamma^5\tau_1\langle\pi\rangle
+\sigma'+i\gamma^5(\tau_1\pi'+\tau_2\langle\pi\rangle\theta+\tau_3\pi'_3)
\end{equation}
to the left-hand side, we substitute Eq.(\ref{KG}) inverted by diagonalizing 
the meson mixing to Eq.(\ref{EOM}). 
Then we perform a one-body reduction (the Wick decomposition) such as 
\begin{equation}
\langle\tilde0\vert T\bar q(y)q(y)q(x)\bar q(x')\vert\tilde0\rangle
\rightarrow\langle\tilde0\vert Tq(x)\bar q(y)\vert\tilde0\rangle
             \langle\tilde0\vert Tq(y)\bar q(x')\vert\tilde0\rangle .
\end{equation}
Note that only the Fock terms appear since the Hartree (mean field) terms have 
already been sorted. Consequently the resulting equation of motion reads 
\begin{equation}
\begin{split}
&(i\Slash{\partial}-m_q-G(\langle\sigma\rangle+i\gamma^5\tau_1\langle\pi\rangle)
  +\frac{\mu_\mathrm{I}}{2}\gamma^0\tau_3)G(x-x')
\\
&=\delta^4(x-x')-\Sigma(x-y)G(y-x') ,
\end{split}
\end{equation}
where $\Sigma(x-y)$ stands for the non-local Fock selfenergy that depends on $G(x-y)$, 
and an integration over $y$ is understood. 
By a Fourier transformation and an isospin decomposition, 
\begin{gather}
A^{ik}=A^0\delta^{ik}+A^3\tau_3^{ik}+A^-\tau_+^{ik}+A^+\tau_-^{ik} ,
\notag \\
\tau_\pm=\frac{1}{\sqrt{2}}(\tau_1\pm i\tau_2) ,
\end{gather}
we obtain a Gor'kov~\cite{Go} type equation, 
\begin{equation}
\left(
\begin{array}{@{\,}cc@{\,}}
 \gamma^0(\omega-h\pm \mu_\mathrm{I}/2)+\Sigma^0\pm\Sigma^3 & -G\langle\pi\rangle i\gamma^5+\sqrt{2}\Sigma^\mp \\
 -G\langle\pi\rangle i\gamma^5+\sqrt{2}\Sigma^\pm & \gamma^0(\omega-h\mp \mu_\mathrm{I}/2)+\Sigma^0\mp\Sigma^3
\end{array}
\right)
\left(
\begin{array}{@{\,}c@{\,}}
 G^0\pm G^3 \\
 \sqrt{2}G^\pm
\end{array}
\right)
=
\left(
\begin{array}{@{\,}c@{\,}}
 1 \\
 0
\end{array}
\right) ,
\label{gortype}
\end{equation}
with $h=\mathbf{\alpha}\cdot\mathbf{k}+\beta(m_q+G\langle\sigma\rangle)$ being the 
free single particle Hamiltonian with the constituent quark mass, 
$M_q=m_q+G\langle\sigma\rangle$. 
This form clearly indicates that the present subject is a pairing problem. 
The upper and lower double signs mean the $u$ and $d$ quark sector, respectively; 
both contain the same information. In the following we take the lower one. 

 In order to solve Eq.(\ref{gortype}) and look into the spatial structure of the 
composite two body system, the pair wave function~\cite{PS} given by the 
Bogoliubov amplitudes is necessary. The route is parallel to the 
non-relativistic case depicted in App.~\ref{appa}. This method was utilized for the 
nucleon pairing in Ref.~\cite{MM_D}. In the present case, $G^0\pm G^3$ corresponds 
to the normal Green function and $\sqrt{2}G^\pm$ does to the anomalous one. 
First we express them in terms of the densities. The relativistic free quark 
field of $i$-th flavor without pairing is expressed as 
\begin{equation}
q_\alpha^i(x)=\frac{1}{\sqrt{V}}\sum_{\mathbf{k}s}
[a_{\mathbf{k}s}^iU_\alpha(\mathbf{k}s)\mathrm{e}^{-ikx}
+b_{\mathbf{k}s}^{i\,\dagger}V_\alpha(\mathbf{k}s)\mathrm{e}^{ikx}]
\label{field}
\end{equation}
with $k^0=E_k\equiv\sqrt{\mathbf{k}^2+M_q^2}$. The number of single particle states 
must be doubled so as to have two energy states mixed by pairing interaction, as done 
by means of the Nambu representation~\cite{Nam} in field theoretical terms. The doubled 
states are diagonalized by means of the Bogoliubov transformation. Then the upper 
half states are regarded as unoccupied quasiparticle states while the lower half ones 
are occupied quasihole states. Therefore the particle states before transformation 
are regarded as superpositions of the quasiparticle with energy $\mathcal{E}_k$ 
and the quasihole with energy $-\mathcal{E}_k$. 
Thus, in the present case, the quark field that defines $G(x-x')$ is thought to be 
expanded in the same form as Eq.(\ref{field}) but with 
\begin{equation}
k^0=\begin{cases}
      +\mathcal{E}_k & \text{particle part with coefficient $u^i$} \\
      -\mathcal{E}_k & \text{hole part with coefficient $v^i$} ,
    \end{cases}
\end{equation}
with the Bogoliubov amplitudes specified below. 
Substituting it to Eq.(\ref{propagator}) and Fourier transformation lead to 
\begin{equation}
\begin{split}
G_{\alpha\beta}^{ij}(\omega,\mathbf{k})
=&\sum_sU_\alpha(\mathbf{k}s)\bar U_\beta(\mathbf{k}s)
\Big(\frac{\langle a_{\mathbf{k}s}^ia_{\mathbf{k}s}^{j\,\dagger}\rangle}
      {\omega-\mathcal{E}_k+i\eta}
+\frac{\langle a_{\mathbf{k}s}^{j\,\dagger}a_{\mathbf{k}s}^i\rangle}
      {\omega+\mathcal{E}_k-i\eta}\Big) \\
+&\sum_sV_\alpha(-\mathbf{k}-s)\bar V_\beta(-\mathbf{k}-s)
\Big(\frac{\langle b_{-\mathbf{k}-s}^{i\,\dagger}b_{-\mathbf{k}-s}^j\rangle}
      {\omega-\mathcal{E}_k+i\eta}
+\frac{\langle b_{-\mathbf{k}-s}^jb_{-\mathbf{k}-s}^{i\,\dagger}\rangle}
      {\omega+\mathcal{E}_k-i\eta}\Big) \\
+&\sum_sU_\alpha(\mathbf{k}s)\bar V_\beta(-\mathbf{k}-s)
\Big(\frac{\langle a_{\mathbf{k}s}^ib_{-\mathbf{k}-s}^j\rangle}
      {\omega-\mathcal{E}_k+i\eta}
+\frac{\langle b_{-\mathbf{k}-s}^ja_{\mathbf{k}s}^i\rangle}
      {\omega+\mathcal{E}_k-i\eta}\Big) \\
+&\sum_sV_\alpha(-\mathbf{k}-s)\bar U_\beta(\mathbf{k}s)
\Big(\frac{\langle b_{-\mathbf{k}-s}^{i\,\dagger}a_{\mathbf{k}s}^{j\,\dagger}\rangle}
      {\omega-\mathcal{E}_k+i\eta}
+\frac{\langle a_{\mathbf{k}s}^{j\,\dagger}b_{-\mathbf{k}-s}^{i\,\dagger}\rangle}
      {\omega+\mathcal{E}_k-i\eta}\Big) .
\end{split}
\end{equation}

Next, the densities such as $\langle aa^\dagger\rangle$ are expressed in terms of 
the Bogoliubov amplitudes by specifying the relevant transformation. 
In general, the exchange of quantum mesonic field produces non-local interactions 
of the type $a^i$--$a^i$, $b^i$--$b^i$, $a^i$--$b^i$, $a^i$--$a^j$, $b^i$--$b^j$, 
and $a^i$--$b^j$ ($i\neq j$).  
Therefore the quasiparticle takes the form of Eq.(\ref{qp}). To be specific, however, 
here we consider $a^i$--$b^j$ that leads to the momentum dependent pionic gap function, 
and $a^i$--$a^i$ and $b^i$--$b^i$ that lead to the Fock mass, among them. 
Then the two types of quasiparticles specified in App.~\ref{appb} 
decouple from each other; $a^i$ and $b^{i\,\dagger}$ in $q^i$ become a constituent 
of different kind of quasiparticles. 
Then the normal and anomalous propagators are given as 
\begin{equation}
\begin{split}
(G^0-G^3)_{\alpha\beta}
&=\Big(\sum_sU_\alpha\bar U_\beta u^2u^{2\,\ast}
     +\sum_sV_\alpha\bar V_\beta v^2v^{2\,\ast}\Big)
\Big(\frac{1}{\omega-\mathcal{E}_k+i\eta}-\frac{1}{\omega+\mathcal{E}_k-i\eta}\Big) , \\
\sqrt{2}G^-_{\alpha\beta}
&=\Big(\sum_sU_\alpha\bar V_\beta u^1v^{2\,\ast}
       +\sum_sV_\alpha\bar U_\beta v^1u^{2\,\ast}\Big)
\Big(\frac{1}{\omega-\mathcal{E}_k+i\eta}-\frac{1}{\omega+\mathcal{E}_k-i\eta}\Big) .
\end{split}
\end{equation}
Note that the expectation values arisen from the commutation relation are already 
subtracted in the backward terms. 

 Substituting these expressions back to Eq.(\ref{gortype}) and taking residues 
at $\omega=\mathcal{E}_k$, finally we obtain a $4\times4$ 
hermitian matrix equation at each $k$, 
\begin{equation}
\begin{split}
&\left(
\begin{array}{@{\,}cccc@{\,}}
 e-E_k-\mu_\mathrm{I}/2-m_2 & 0 & -\pi & 0 \\
 0 & e+E_k-\mu_\mathrm{I}/2-\tilde m_2 & 0 & -\tilde\pi \\
 -\pi & 0 & e+E_k+\mu_\mathrm{I}/2-\tilde m_1 & 0 \\
 0 & -\tilde\pi & 0 & e-E_k+\mu_\mathrm{I}/2-m_1
\end{array}
\right)
\\
&\times
\left(
\begin{array}{@{\,}c@{\,}}
 A \\
 B \\
 C \\
 D 
\end{array}
\right)
=0 .
\end{split}
\label{mat}
\end{equation}
Here the eigenenergy is denoted by $e$ since both the quasiparticle and 
quasihole solutions are obtained from this, and use has been made of 
\begin{equation}
hU=E_kU,\quad hV=-E_kV .
\end{equation}
The real Bogoliubov amplitudes are defined as 
\begin{gather}
A=u^2=\langle\tilde0\vert a_d\eta^\dagger\vert\tilde0\rangle,\quad
B=v^2=\langle\tilde0\vert b_{-d}^\dagger\eta^\dagger\vert\tilde0\rangle,
\notag \\
C=-iv^1=-i\langle\tilde0\vert b_{-u}^\dagger\eta^\dagger\vert\tilde0\rangle,\quad
D=-iu^1=-i\langle\tilde0\vert a_u\eta^\dagger\vert\tilde0\rangle,
\end{gather}
and all quantities appearing in Eq.(\ref{mat}) are real. 
Among them, 
\begin{equation}
\begin{split}
\pi(k)&=-i\bar U(k)(-G\langle\pi\rangle i\gamma^5+\sqrt{2}\Sigma^+)V(k) ,\\
\tilde\pi(k)&=-i\bar V(k)(-G\langle\pi\rangle i\gamma^5+\sqrt{2}\Sigma^+)U(k) ,
\end{split}
\label{bf}
\end{equation}
represent the momentum dependent pionic gap functions for the $d\bar u$ and 
$u\bar d$ condensation, respectively, while 
\begin{equation}
\begin{split}
&m_2(k)=-\bar U(k)(\Sigma^0-\Sigma^3)U(k) , \\
&\tilde m_2(k)=-\bar V(k)(\Sigma^0-\Sigma^3)V(k) , \\
&\tilde m_1(k)=-\bar V(k)(\Sigma^0+\Sigma^3)V(k) , \\
&m_1(k)=-\bar U(k)(\Sigma^0+\Sigma^3)U(k) 
\end{split}
\end{equation}
do the Fock masses. 
The first term in each equation in Eq.(\ref{bf}) stems from the momentum 
independent pion condensation 
$\langle\pi\rangle$ of the meson system, which produces a strong momentum 
dependence, $\bar U\gamma^5V=M_q/E_k$, and the second one from the non-local 
Fock selfenergy. This type of $4\times4$ matrix equation appears also in the 
cases of the relativistic 1 flavor pairing including the Dirac sea~\cite{MM_D} 
and the non-relativistic 2 flavor pairing~\cite{CCJ}. 
Since the Fock selfenergy at a momentum $k$ is a function of $A(k')$--$D(k')$, 
the equations for all momenta are coupled. 
Actually, when evaluating each matrix element of $\Sigma$, a 4-momentum integration 
is necessary. For the energy integration among them, we make an instantaneous 
approximation, that is, energy transfer$\rightarrow0$ as in previous 
works~\cite{MM_D,MM_q,AHI}. As for the remaining 3-momentum integration, the BCS 
type calculation needs a cutoff in general. In the present case it is thought to be 
around the typical hadronic scale. Therefore we adopt that for the standard NJL 
model for simplicity. 
Solving the coupled equations selfconsistently determines all the physical 
quantities: The Bogoliubov amplitudes, quasiparticle energies, and the mass and 
gap functions at each $\mu_\mathrm{I}$. Then the pair wave functions and the coherence 
length are calculated from them. 

Now we proceed to numerical calculations. Parameters used are the current quark mass 
$m_q=$ 0.0055 GeV, the momentum cutoff $\Lambda=$ 0.63 GeV, the pion 
decay constant $f_\pi=$ 0.093 GeV, the pion mass $m_\pi=$ 0.138 GeV, the potential 
parameter in the linear sigma model $\lambda=$ 4.5, and the quark--meson coupling 
$G=$ 3.3. The momentum space $0\le k \le \Lambda$ is divided to 100 equi-intervals 
for the coupled Newton method. 
Calculations are done for $\mu_\mathrm{I}<0$ where the $d\bar u$ 
condensation dominates. The results depend on the parameters quantitatively but 
the qualitative behavior is robust; this will be confirmed later with respect to 
the behavior of the coherence length, which is of direct physical relevance. 

\begin{figure}[htbp]
 \includegraphics[width=5cm,angle=-90]{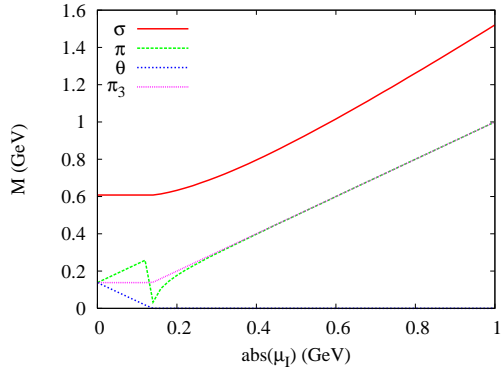}
 \caption{(Color online) Meson masses given by the linear sigma model with the 
approximation described in the text. Note that $\pi$ and $\theta$ correspond to 
$\pi_1$ and $\pi_2$, respectively, at $|\mu_\mathrm{I}|<m_\pi$.}
 \label{fig1}
\end{figure}

 First, we check the meson masses under the present approximation in Fig.~\ref{fig1}. 
The cusp just after the transition $|\mu_\mathrm{I}|=m_\pi$ is brought about by the 
neglect of the Coriolis coupling term in Eq.(\ref{KG}). 
Definitely, the eigenvalues of the $2\times2$ diagonalization after that in the polar 
coordinate representation are 
\begin{equation}
M^2=\frac{1}{2}\Big(2\lambda^2(\langle\sigma\rangle^2+\langle\pi\rangle^2)+\mu_\mathrm{I}^2
          \pm\sqrt{[2\lambda^2(\langle\sigma\rangle^2-\langle\pi\rangle^2)+\mu_\mathrm{I}^2]^2
          +16\lambda^4\langle\sigma\rangle^2\langle\pi\rangle^2}\Big) ,
\label{mymass}
\end{equation}
while two non-zero eigenvalues of the $3\times3$ diagonalization in the Cartesian coordinate 
representation~\cite{HJZ} are 
\begin{equation}
M^2=\frac{1}{2}\Big(2\lambda^2(\langle\sigma\rangle^2+\langle\pi\rangle^2)+5\mu_\mathrm{I}^2
          \pm\sqrt{[2\lambda^2(\langle\sigma\rangle^2-\langle\pi\rangle^2)-3\mu_\mathrm{I}^2]^2
          +16\lambda^4\langle\sigma\rangle^2\langle\pi\rangle^2}\Big) .
\end{equation}
The present result given by Eq.(\ref{mymass}), the lower one of which tends to 0 when 
$\langle\pi\rangle$ approaches 0, is not consistent with the one obtained in 
the frame of the chiral perturbation~\cite{LV2}, but this difference is a trade-off for obtaining 
the meson mixing. 
Practically, its influence is limited to just after the transition. 

\begin{figure}[htbp]
 \includegraphics[width=5cm,angle=-90]{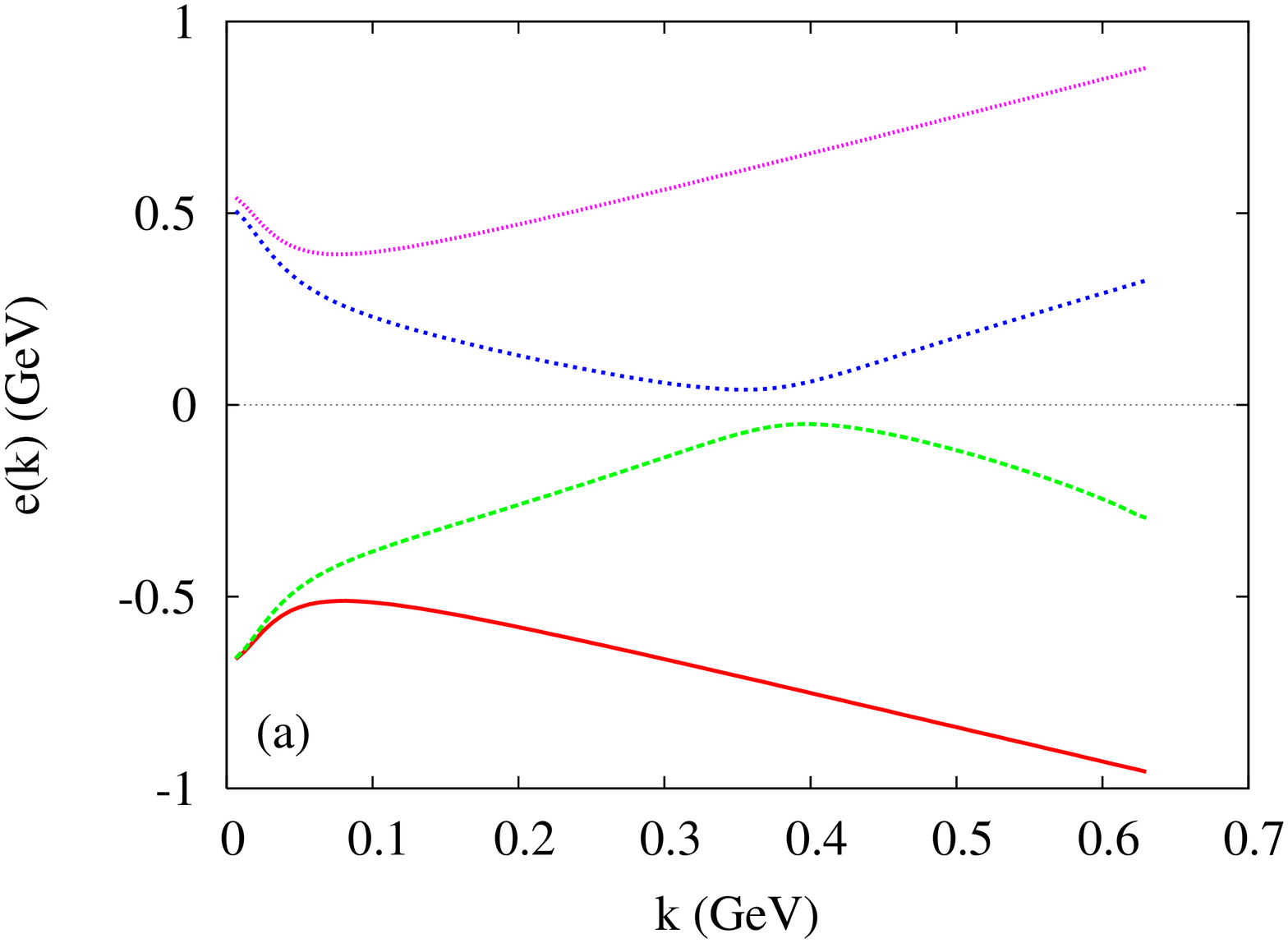}
 \includegraphics[width=5cm,angle=-90]{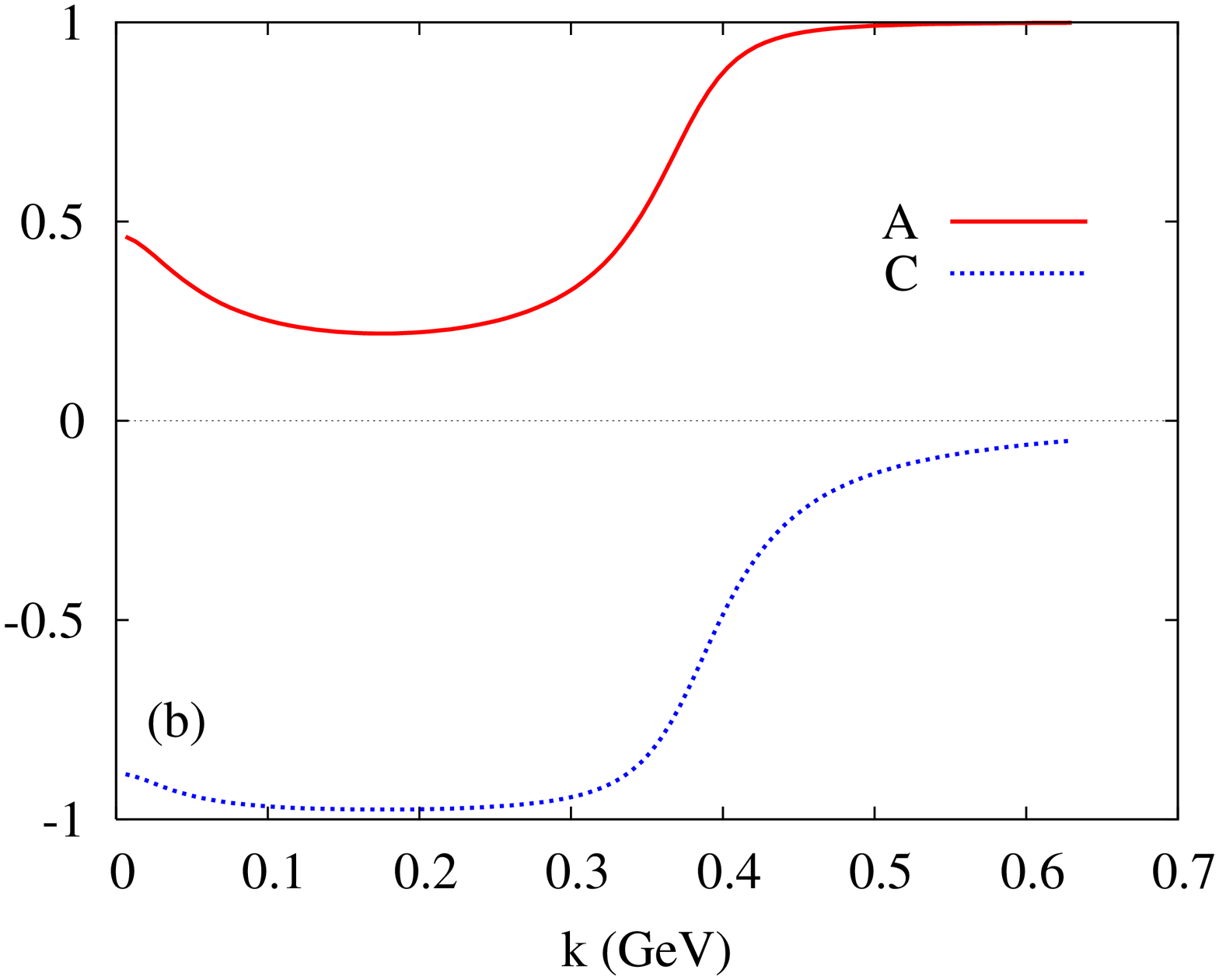}
 \includegraphics[width=5cm,angle=-90]{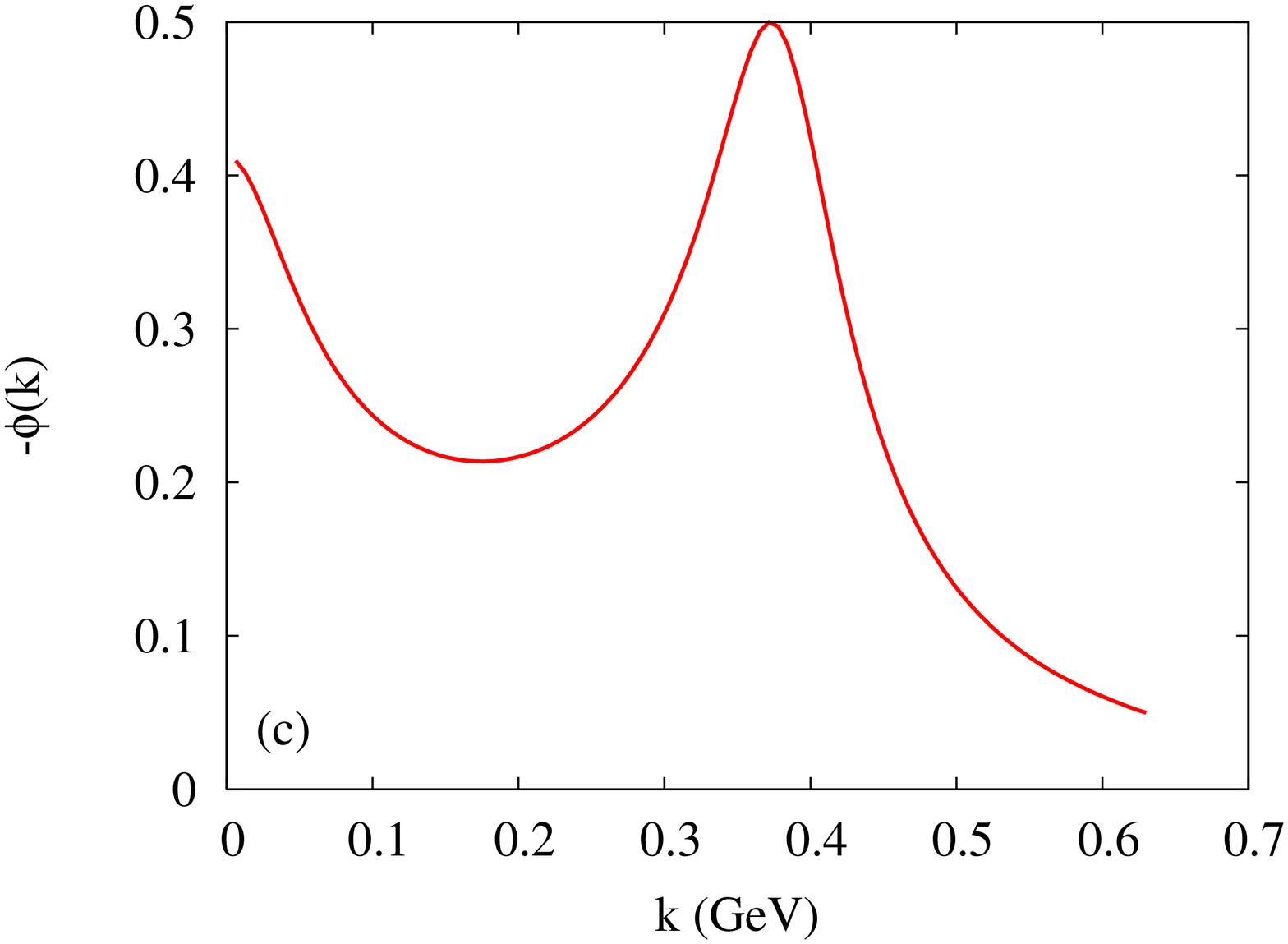}
 \includegraphics[width=5cm,angle=-90]{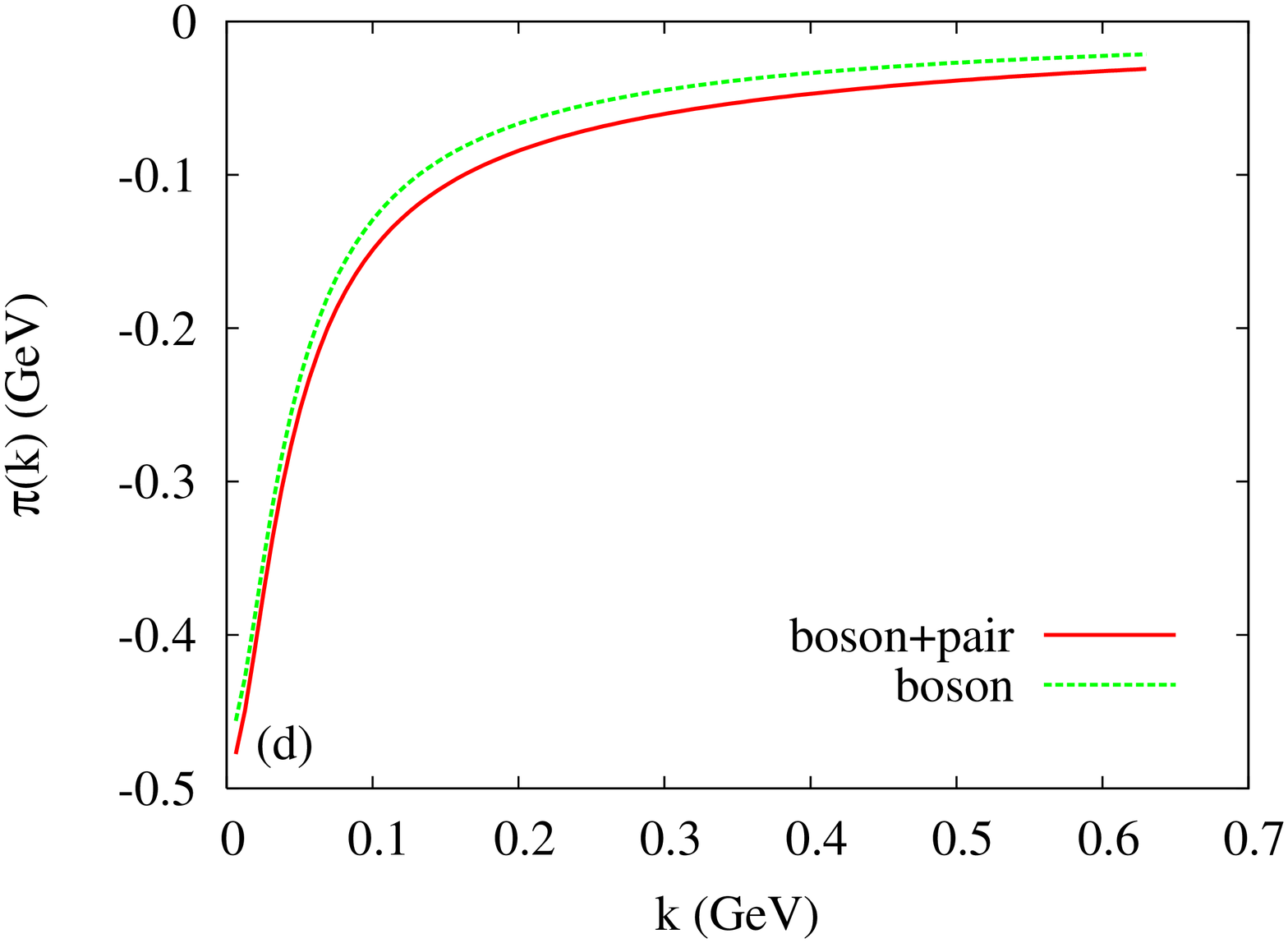}
 \caption{(Color online) Momentum dependence of various quantities at 
$\mu_\mathrm{I}= -0.5$ GeV: (a) the quasiparticle energies, (b) the Bogoliubov amplitudes, 
(c) the pair wave function, and (d) the gap function. Note that (b) -- (d) are associated 
with the third (from the bottom) solution in (a).}
 \label{fig2}
\end{figure}

 Figure~\ref{fig2} shows the results at $|\mu_\mathrm{I}|=$ 0.5 GeV $\gg m_\pi$. 
Figure~\ref{fig2}~(a) is the quasiparticle energy diagram as a function of the 
relative momentum $k$ (dispersion relation). Its unperturbed structure is quite 
simple: The positive and negative energy $u$ ($d$) quark levels with $\pm E_k$ 
are shifted upward (downward) by $|\mu_\mathrm{I}|/2$. Then, the negative energy 
$u$, that is the hole state of the $\bar u$, and the positive energy $d$ interact 
around the Fermi surface. This means the $d\bar u$ pairing. Hereafter we name 
these quasiparticle (hole) levels the first, second, third and fourth, from the 
bottom. The third level, the lower quasiparticle, is the main interest in the 
following discussion. This lower quasiparticle consists only of $A$ and $C$. 
In the usual pairing problem, for example in the case of Ref.~\cite{MM_D}, this type of 
$2\times2$ equation can be cast into the form of the gap equation. In the present 
case, however, $\pi(k)$ is represented as a function of $A(k')$ and $C(k')$ as 
\begin{gather}
\pi(k)=-\frac{1}{2}\sum_{k'}^\Lambda(v(k,k')2A(k')C(k')+v'(k,k')(A^2(k')-C^2(k'))), 
\notag \\
2A(k')C(k')=\frac{\pi(k')}{e(k')-\frac{m_1(k')+\tilde m_2(k')}{2}}, 
\notag \\
A^2(k')-C^2(k')=\frac{E_{k'}+\frac{\mu_\mathrm{I}}{2}+\frac{m_1(k')-\tilde m_2(k')}{2}}
                     {e(k')-\frac{m_1(k')+\tilde m_2(k')}{2}}. 
\label{pwf}
\end{gather}
Therefore the $v'$ term due to the $\sigma$--$\pi$ mixing prevents one from casting 
Eq.(\ref{mat}) into the form of the gap equation. 
Nevertheless, the notion of the pair wave function~\cite{PS} is useful for looking into 
the physical contents since $A^2-C^2$ is small around the Fermi surface. 
Figure~\ref{fig2}~(b) shows the Bogoliubov amplitudes $A$ and 
$C$. Aside from the bump around $k=0$ mentioned below, the hole character changes gradually 
to the particle character around the Fermi surface as the usual Cooper pairing. 
This leads to the peak in the pair wave function $\phi(k)=A(k)C(k)$ (see Eq.(\ref{pwf})) 
shown in Fig.~\ref{fig2}~(c). The bump around $k=0$ is a novel feature of the present case; 
this is brought about by the mesonic contribution $\langle\pi\rangle$ to the gap 
function $\pi(k)$ (see Eq.(\ref{bf})) as shown in Fig.~\ref{fig2}~(d). 
In this gap function, the mesonic and the Cooper pair components are 
comparable around the Fermi surface, whereas the former is dominant around $k=0$ because of 
the $k$ dependence $\propto M_q/E_k$. 

\begin{figure}[htbp]
 \includegraphics[width=5cm,angle=-90]{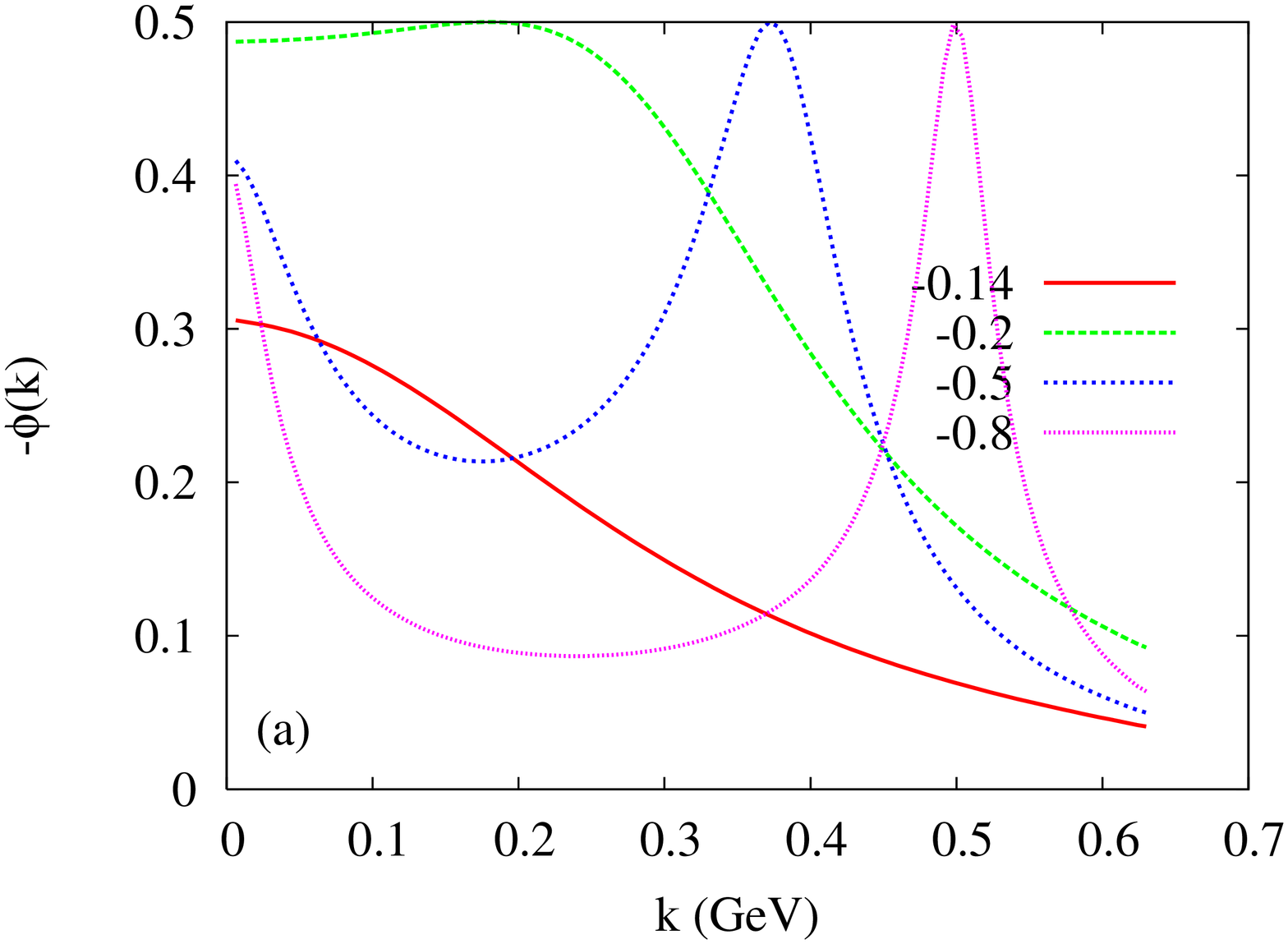}
 \includegraphics[width=5cm,angle=-90]{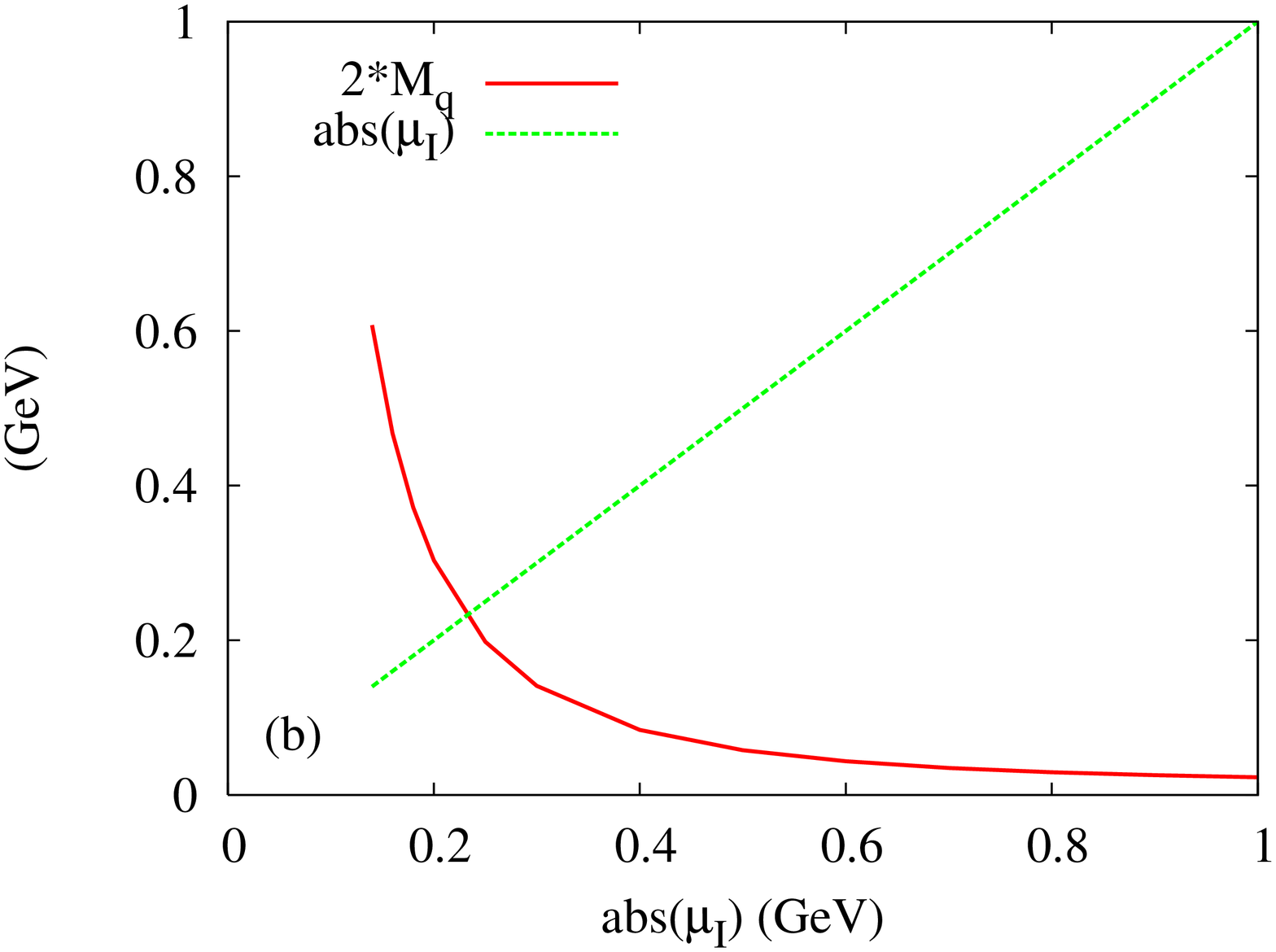}
 \includegraphics[width=5cm,angle=-90]{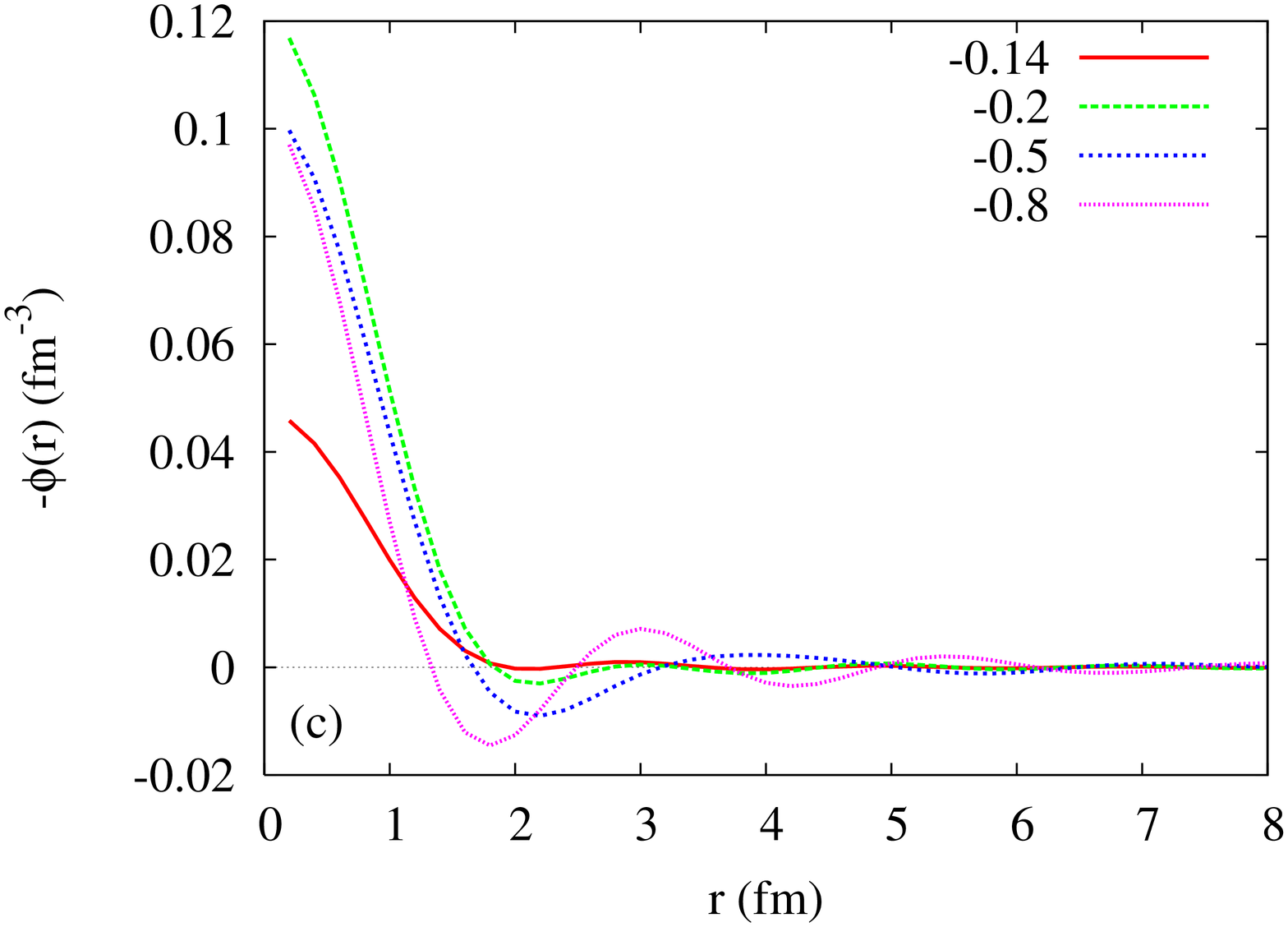}
 \includegraphics[width=5cm,angle=-90]{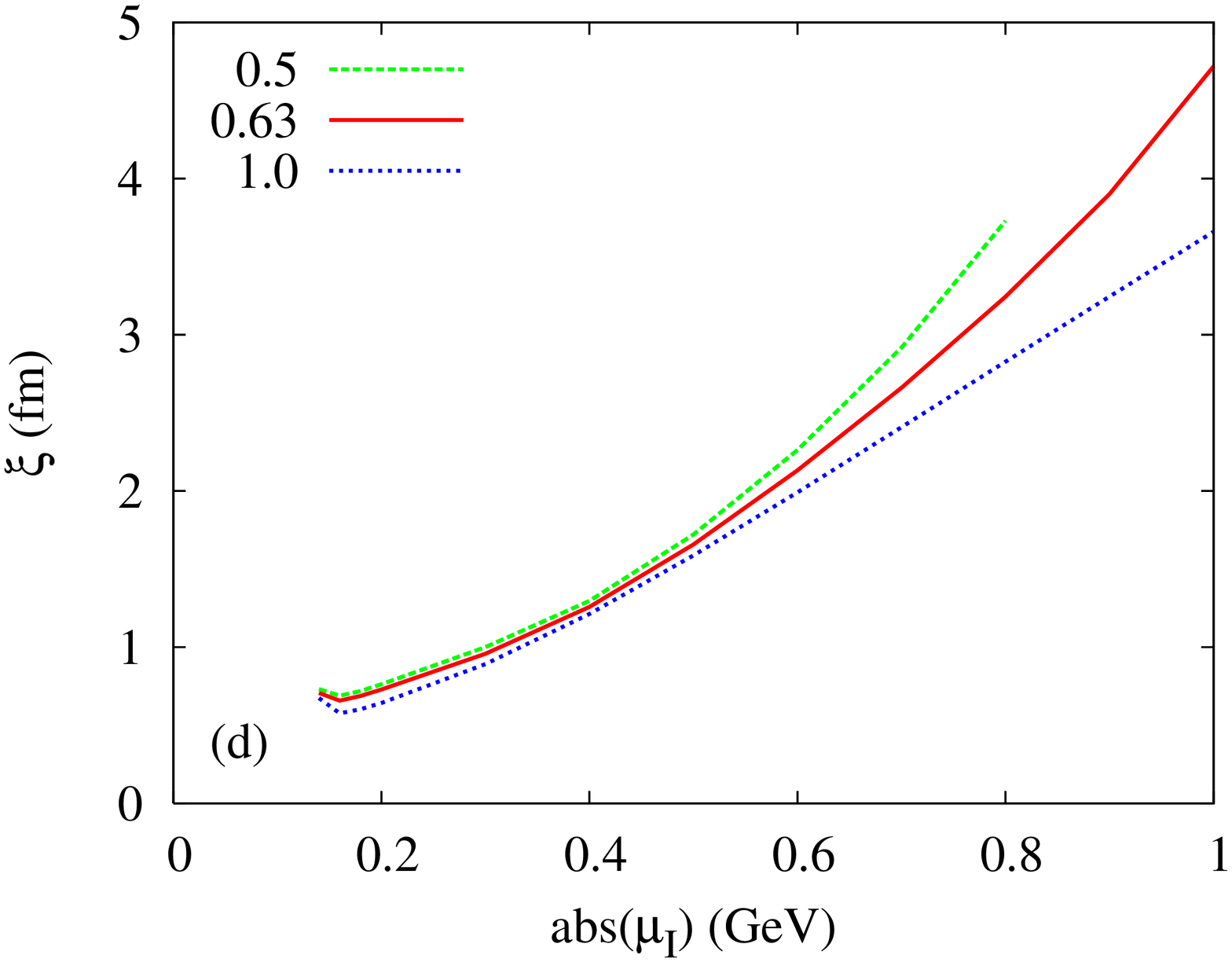}
 \includegraphics[width=5cm,angle=-90]{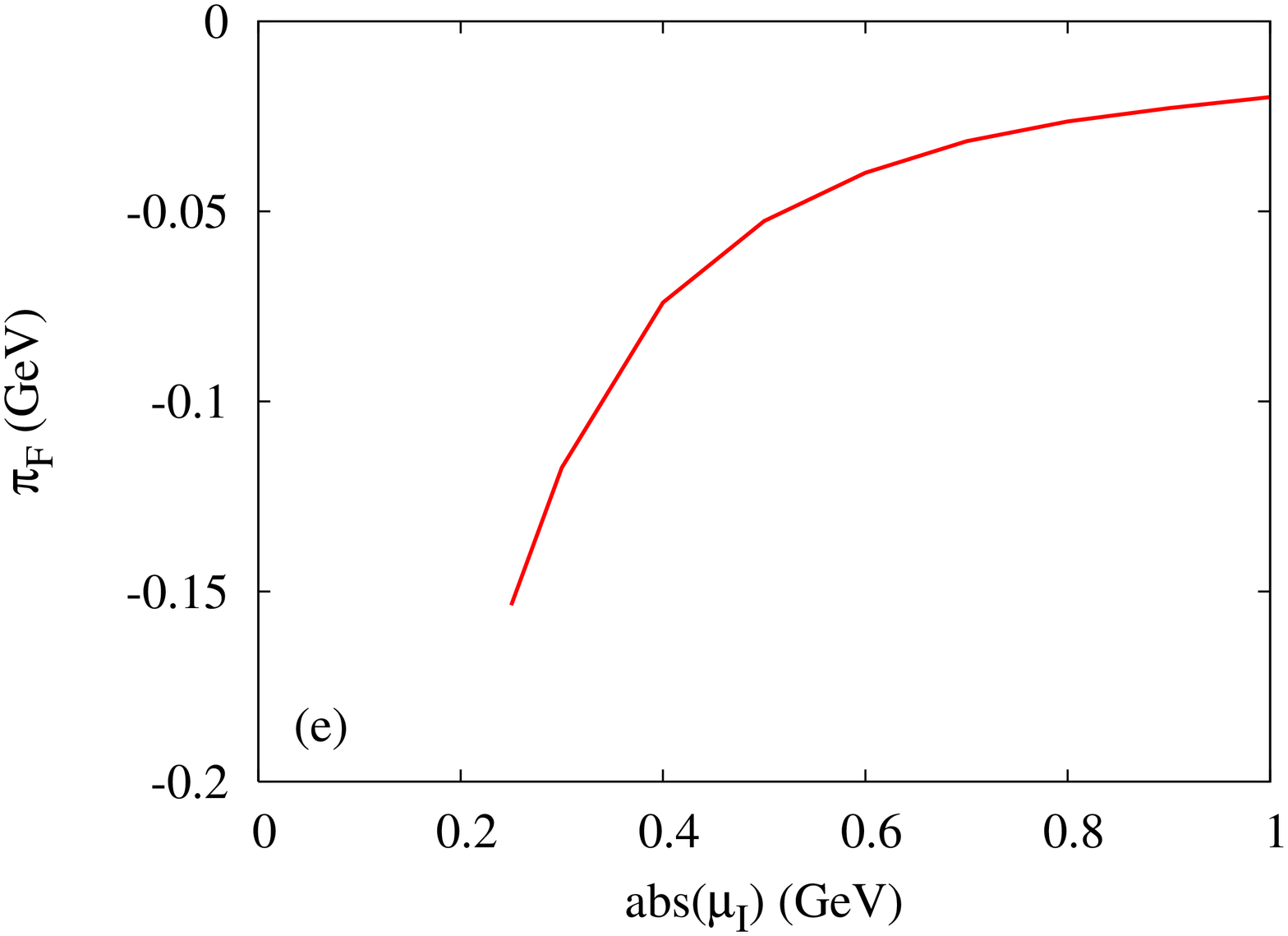}
 \caption{(Color online) Isospin chemical potential dependence of various quantities: 
(a) the $k$ space pair wave function, (b) the twice of the constituent quark mass, 
(c) the $r$ space pair wave function, (d) the coherence length, and (e) the gap at 
the Fermi surface. (d) also contains the cutoff dependence.}
 \label{fig3}
\end{figure}

 Figure~\ref{fig3} shows the $\mu_\mathrm{I}$ dependence of various quantities. 
Figure~\ref{fig3}~(a) shows the pair wave functions at several $\mu_\mathrm{I}$s as 
functions of the momentum. This shows that, leaving room for possible error related 
to the discussion about Fig.~\ref{fig1}, at low $|\mu_\mathrm{I}|$ the peak due to 
the Cooper pairing can not be seen. Actually, $q$ and $\bar q$ are bound to each 
other for $|\mu_\mathrm{I}|<2M_q$ as shown in Fig.~\ref{fig3}~(b). Thus, we can 
conclude that the pionic condensation has a mixed character: Purely bosonic just 
after the appearance of the condensation, then the Cooper pairing gradually grows 
as $|\mu_\mathrm{I}|$ increases with retaining significant bosonic component. 
To look into the spatial structure of Cooper pairs more closely, we Fourier transform 
$\phi(k)$ as 
\begin{equation}
\phi(r)=\frac{1}{2\pi^2}\int_0^\Lambda\phi(k)j_0(kr)k^2dk.
\end{equation}
The results for several $\mu_\mathrm{I}$s are shown in Fig.~\ref{fig3}~(c) as 
functions of the relative distance. 
Obviously those for higher $|\mu_\mathrm{I}|$ wave till longer distance. 
Figure~\ref{fig3}~(d) graphs the coherence length, 
\begin{equation}
\xi=\left(\frac{\int_0^\Lambda\vert\frac{d\phi}{dk}\vert^2k^2dk}
               {\int_0^\Lambda\vert\phi\vert^2k^2dk}\right)^{1/2},
\end{equation}
and \ref{fig3}~(e) the gap at the Fermi surface as functions of $\mu_\mathrm{I}$. 
The obtained coherence length at low $|\mu_\mathrm{I}|$ is consistent with the 
value obtained by an analysis of the $\pi$-$\pi$ scattering, 
$\langle r^2\rangle_S^\pi=0.61\pm0.04$ fm$^2$~\cite{CGL}. 
In relation to heavy ion collisions, this value is very close to the typical 
inter-pion distance $d$ at the freeze-out: An example of numbers, the charged 
particle multiplicity $N_\mathrm{c}=555$~\cite{Ba} and the source size 
$V=(\mathrm{6.48 fm})^3$~\cite{BF}, and the fact that the pion is the most abundant, 
lead to $d\agt(V/N_\mathrm{c})^{1/3}=$ 0.79 fm. The picture of a gas of bound mesons 
may apply to $\xi<d$ while that of a liquid (see also Ref.~\cite{AS}) of Cooper 
pairs would be appropriate for $\xi>d$ although the latter realizes at rather 
high $|\mu_\mathrm{I}|$. 
Figure~\ref{fig3} clearly indicates that the Cooper pairing becomes weakly coupled as 
$|\mu_\mathrm{I}|$ increases. 
Comparing these figures with corresponding ones in Ref.~\cite{MM_q}, 
one can see that the Cooper pairing part of the present case is more weakly coupled 
than the case of color superconductivity, as represented by the narrower peak in 
$\phi(k)$ and longer spatial extent. Figure~\ref{fig3} (d) also shows the cutoff 
dependence; the dependence is weak. 

 At higher $|\mu_\mathrm{I}|$, in the present calculation $|\mu_\mathrm{I}|\ge$ 0.8 GeV, 
a gapless pairing ($e<0$) takes place. The gapless dispersion is known to occur 
in the case of pairing between particles with different masses~\cite{WY}. 
In the present case, the Fock term produces the difference in the mass 
(see the denominator in Eq.(\ref{pwf})). 

 Finally we look into the character of the fourth level, the higher quasiparticle, 
that corresponds to the Dirac sea pairing in Ref.~\cite{MM_D}. This level is of almost 
pure $u$ quark particle character ($D(k)\simeq1$) for $k\agt$ 0.1 GeV; but the $\bar d$ 
component strongly mixes around $k=0$ because of two reasons: 1) $\langle\pi\rangle$ 
equally contributes to $\tilde\pi(k)$ and $\pi(k)$ (but with the opposite sign), 
and 2) the unperturbed energy difference between $u$ and $\bar d$ is the same as 
that between $\bar u$ and $d$ at $k=0$. 

 To summarize, we have studied the momentum dependence of the pionic gap function 
$\pi(k)$ that determines the spatial structure of the condensation by adopting the 
linear sigma model as an inter-quark interaction at finite 
isospin chemical potential as a first step towards the study of the asymmetric matter 
in the real world. 
Although confinement is not taken into account in the present study, 
the character of the condensation is bosonic at low $|\mu_\mathrm{I}|$, then 
the Cooper pairing gradually grows as $|\mu_\mathrm{I}|$ increases. 
This $q$--$\bar q$ pairing is weaker than the $q$--$q$ pairing of the case of color 
superconductivity. 
The spatial structure (wave function) of the composite pionic system is expected 
to be measured in lattice QCD simulations as well as the $\mu_\mathrm{I}$ 
dependence of the magnitude of the condensation as signatures of the BEC--BCS 
crossover. The spatial structure may affect the description of pions created in heavy 
ion collisions. 

\appendix
\section{The Gor'kov formalism}
\label{appa}

 Gor'kov~\cite{Go} first proposed a field theoretical method to describe the 
pairing problem. In addition to the normal Green function $G(x-x')$, the anomalous 
Green function $F^\dagger(x-x')$ of $\langle T(\psi^\dagger\psi^\dagger)\rangle$ 
type is introduced there. The equation of motion of their Fourier transforms is 
given by 
\begin{equation}
\left(
\begin{array}{@{\,}cc@{\,}}
 \omega-\xi_k & -i\Delta  \\
 i\Delta & \omega+\xi_k 
\end{array}
\right)
\left(
\begin{array}{@{\,}c@{\,}}
 G(\omega,\mathbf{k}) \\
 F^\dagger(\omega,\mathbf{k}) 
\end{array}
\right)
=
\left(
\begin{array}{@{\,}c@{\,}}
 1 \\
 0 
\end{array}
\right)
 ,
\label{eqa1}
\end{equation}
where $\xi_k$ and $\Delta$ are the single particle energy measured from the 
Fermi surface and the momentum independent pairing gap, respectively. 
Its solution is 
\begin{gather}
G(\omega,\mathbf{k})=\frac{u_k^2}{\omega-\mathcal{E}_k}
                    +\frac{v_k^2}{\omega+\mathcal{E}_k} ,
\notag \\
F^\dagger(\omega,\mathbf{k})=\frac{-i\Delta/2\mathcal{E}_k}{\omega-\mathcal{E}_k}
                            +\frac{ i\Delta/2\mathcal{E}_k}{\omega+\mathcal{E}_k} ,
\notag \\
u_k^2=\frac{1}{2}\Big(1+\frac{\xi_k}{\mathcal{E}_k}\Big) ,\quad 
v_k^2=\frac{1}{2}\Big(1-\frac{\xi_k}{\mathcal{E}_k}\Big) , 
\notag \\
u_k^2v_k^2=\Big(\frac{\Delta}{2\mathcal{E}_k}\Big)^2, \quad 
\mathcal{E}_k=\sqrt{\xi_k^2+\Delta^2} .
\end{gather}
Substituting them back to Eq.(\ref{eqa1}) gives 
\begin{equation}
\left(
\begin{array}{@{\,}cc@{\,}}
 \frac{[(\omega-\xi_k)u_k-\Delta v_k]u_k}{\omega-\mathcal{E}_k} 
+\frac{[(\omega-\xi_k)v_k+\Delta u_k]v_k}{\omega+\mathcal{E}_k} \\ 
 i\frac{[\Delta u_k-(\omega+\xi_k)v_k]u_k}{\omega-\mathcal{E}_k} 
+i\frac{[\Delta v_k+(\omega+\xi_k)u_k]v_k}{\omega+\mathcal{E}_k} 
\end{array}
\right)
=
\left(
\begin{array}{@{\,}c@{\,}}
 1 \\
 0 
\end{array}
\right)
 .
\end{equation}
The residues at $\omega=\mathcal{E}_k$ (quasiparticle) lead to 
\begin{equation}
\left(
\begin{array}{@{\,}cc@{\,}}
 \xi_k & \Delta  \\
 \Delta & -\xi_k 
\end{array}
\right)
\left(
\begin{array}{@{\,}c@{\,}}
 u_k \\
 v_k 
\end{array}
\right)
=\mathcal{E}_k
\left(
\begin{array}{@{\,}c@{\,}}
 u_k \\
 v_k 
\end{array}
\right)
 .
\end{equation}
Those at $\omega=-\mathcal{E}_k$ (quasihole) lead to the same equation. 
Therefore the equation for the Green functions and that for the Bogoliubov 
amplitudes are equivalent. 

\section{The Bogoliubov transformation}
\label{appb}

 Replacing the spin $\sigma=\uparrow/\downarrow$ and $a_{-\mathbf{k}}$ in the 
non-relativistic pairing problem by the isospin $u/d$ and $b_{-\mathbf{k}}$, 
respectively, we obtain two Bogoliubov transformations relevant to the present 
case, 
\begin{equation}
\begin{split}
&\left(
\begin{array}{@{\,}c@{\,}}
 a_u \\
 b_{-d}^\dagger 
\end{array}
\right)
=
\left(
\begin{array}{@{\,}cc@{\,}}
 u^1 & -v^2  \\
 v^2 & -u^1 
\end{array}
\right)
\left(
\begin{array}{@{\,}c@{\,}}
 \eta_u \\
 \eta_{-d}^\dagger 
\end{array}
\right)
 , \\
&\left(
\begin{array}{@{\,}c@{\,}}
 a_d \\
 b_{-u}^\dagger 
\end{array}
\right)
=
\left(
\begin{array}{@{\,}cc@{\,}}
 u^2 & v^1  \\
 v^1 & u^2 
\end{array}
\right)
\left(
\begin{array}{@{\,}c@{\,}}
 \eta_d \\
 \eta_{-u}^\dagger 
\end{array}
\right)
 , 
\end{split}
\end{equation}
at each momentum and spin. 

 Since there is $i\gamma^5\overrightarrow\tau$ between two flavors, here we take 
$u^1$ and $v^1$ are imaginary, $u^2$ and $v^2$ are real. Then the two types of 
quasiparticle, $\eta_u^\dagger$ and $\eta_d^\dagger$, can be represented 
collectively as 
\begin{equation}
\eta^\dagger=\sum_{i=1}^2(u^ia_i^\dagger+v^ib_{-i}) ,
\label{qp}
\end{equation}
where $i=1/2$ correspond to $u/d$.


\begin{thebibliography}{46}
\expandafter\ifx\csname natexlab\endcsname\relax\def\natexlab#1{#1}\fi
\expandafter\ifx\csname bibnamefont\endcsname\relax
  \def\bibnamefont#1{#1}\fi
\expandafter\ifx\csname bibfnamefont\endcsname\relax
  \def\bibfnamefont#1{#1}\fi
\expandafter\ifx\csname citenamefont\endcsname\relax
  \def\citenamefont#1{#1}\fi
\expandafter\ifx\csname url\endcsname\relax
  \def\url#1{\texttt{#1}}\fi
\expandafter\ifx\csname urlprefix\endcsname\relax\def\urlprefix{URL }\fi
\providecommand{\bibinfo}[2]{#2}
\providecommand{\eprint}[2][]{\url{#2}}

\bibitem[{\citenamefont{{J. B. Kogut} and {D. K.
  Sinclair}}(2002{\natexlab{a}})}]{KS}
\bibinfo{author}{\bibnamefont{{J. B. Kogut}}} \bibnamefont{and}
  \bibinfo{author}{\bibnamefont{{D. K. Sinclair}}}, \bibinfo{journal}{Phys.\
  Rev.\ D} \textbf{\bibinfo{volume}{66}}, \bibinfo{pages}{034505}
  (\bibinfo{year}{2002}{\natexlab{a}}).

\bibitem[{\citenamefont{{M. Frank} et~al.}(2003)\citenamefont{{M. Frank}, {M.
  Buballa}, and {M. Oertel}}}]{FBO}
\bibinfo{author}{\bibnamefont{{M. Frank}}}, \bibinfo{author}{\bibnamefont{{M.
  Buballa}}}, \bibnamefont{and} \bibinfo{author}{\bibnamefont{{M. Oertel}}},
  \bibinfo{journal}{Phys.\ Lett.\ B} \textbf{\bibinfo{volume}{562}},
  \bibinfo{pages}{221} (\bibinfo{year}{2003}).

\bibitem[{\citenamefont{{Y. Nishida}}(2004)}]{Ni}
\bibinfo{author}{\bibnamefont{{Y. Nishida}}}, \bibinfo{journal}{Phys.\ Rev.\ D}
  \textbf{\bibinfo{volume}{69}}, \bibinfo{pages}{094501}
  (\bibinfo{year}{2004}).

\bibitem[{\citenamefont{{A. Barducci} et~al.}(2004)\citenamefont{{A. Barducci},
  {R. Casalbuoni}, {G. Pettini}, and {L.Ravagli}}}]{BCPR}
\bibinfo{author}{\bibnamefont{{A. Barducci}}},
  \bibinfo{author}{\bibnamefont{{R. Casalbuoni}}},
  \bibinfo{author}{\bibnamefont{{G. Pettini}}}, \bibnamefont{and}
  \bibinfo{author}{\bibnamefont{{L.Ravagli}}}, \bibinfo{journal}{Phys.\ Rev.\
  D} \textbf{\bibinfo{volume}{69}}, \bibinfo{pages}{096004}
  (\bibinfo{year}{2004}).

\bibitem[{\citenamefont{{M. Loewe} and {C. Villavicencio}}(2005)}]{LV}
\bibinfo{author}{\bibnamefont{{M. Loewe}}} \bibnamefont{and}
  \bibinfo{author}{\bibnamefont{{C. Villavicencio}}}, \bibinfo{journal}{Phys.\
  Rev.\ D} \textbf{\bibinfo{volume}{71}}, \bibinfo{pages}{094001}
  (\bibinfo{year}{2005}).

\bibitem[{\citenamefont{{S. Mukherjee} et~al.}(2007)\citenamefont{{S.
  Mukherjee}, {M. G. Mustafa}, and {R. Ray}}}]{MMR}
\bibinfo{author}{\bibnamefont{{S. Mukherjee}}},
  \bibinfo{author}{\bibnamefont{{M. G. Mustafa}}}, \bibnamefont{and}
  \bibinfo{author}{\bibnamefont{{R. Ray}}}, \bibinfo{journal}{Phys.\ Rev.\ D}
  \textbf{\bibinfo{volume}{75}}, \bibinfo{pages}{094015}
  (\bibinfo{year}{2007}).

\bibitem[{\citenamefont{{Z. Zhang} and {Y.-X. Liu}}(2007)}]{ZL}
\bibinfo{author}{\bibnamefont{{Z. Zhang}}} \bibnamefont{and}
  \bibinfo{author}{\bibnamefont{{Y.-X. Liu}}}, \bibinfo{journal}{Phys.\ Rev.\
  C} \textbf{\bibinfo{volume}{75}}, \bibinfo{pages}{064910}
  (\bibinfo{year}{2007}).

\bibitem[{\citenamefont{{J. O. Andersen} and {T. Brauner}}(2008)}]{AB}
\bibinfo{author}{\bibnamefont{{J. O. Andersen}}} \bibnamefont{and}
  \bibinfo{author}{\bibnamefont{{T. Brauner}}}, \bibinfo{journal}{Phys.\ Rev.\
  D} \textbf{\bibinfo{volume}{78}}, \bibinfo{pages}{014030}
  (\bibinfo{year}{2008}).

\bibitem[{\citenamefont{{D. K. Campbell} et~al.}(1975)\citenamefont{{D. K.
  Campbell}, {R. F. Dashen}, and {J. T. Manassah}}}]{CDM}
\bibinfo{author}{\bibnamefont{{D. K. Campbell}}},
  \bibinfo{author}{\bibnamefont{{R. F. Dashen}}}, \bibnamefont{and}
  \bibinfo{author}{\bibnamefont{{J. T. Manassah}}}, \bibinfo{journal}{Phys.\
  Rev.\ D} \textbf{\bibinfo{volume}{12}}, \bibinfo{pages}{979}
  (\bibinfo{year}{1975}).

\bibitem[{\citenamefont{{D. T. Son} and {M. A. Stephanov}}(2001)}]{SS}
\bibinfo{author}{\bibnamefont{{D. T. Son}}} \bibnamefont{and}
  \bibinfo{author}{\bibnamefont{{M. A. Stephanov}}}, \bibinfo{journal}{Phys.\
  Rev.\ Lett.} \textbf{\bibinfo{volume}{86}}, \bibinfo{pages}{592}
  (\bibinfo{year}{2001}).

\bibitem[{\citenamefont{{A. J. Leggett}}(1980)}]{Le}
\bibinfo{author}{\bibnamefont{{A. J. Leggett}}}, \emph{\bibinfo{title}{Modern
  Trends in the Theory of Condensed Matter}}
  (\bibinfo{publisher}{Springer-Verlag}, \bibinfo{year}{1980}),
  p.~\bibinfo{pages}{13}.

\bibitem[{\citenamefont{{P. Nozi{\`{e}}res} and {S. Schmitt-Rink}}(1985)}]{NS}
\bibinfo{author}{\bibnamefont{{P. Nozi{\`{e}}res}}} \bibnamefont{and}
  \bibinfo{author}{\bibnamefont{{S. Schmitt-Rink}}}, \bibinfo{journal}{J.\
  Low.\ Temp.\ Phys.} \textbf{\bibinfo{volume}{59}}, \bibinfo{pages}{195}
  (\bibinfo{year}{1985}).

\bibitem[{\citenamefont{{H. E. Haber} and {H. A. Weldon}}(1982)}]{HW}
\bibinfo{author}{\bibnamefont{{H. E. Haber}}} \bibnamefont{and}
  \bibinfo{author}{\bibnamefont{{H. A. Weldon}}}, \bibinfo{journal}{Phys.\
  Rev.\ D} \textbf{\bibinfo{volume}{25}}, \bibinfo{pages}{502}
  (\bibinfo{year}{1982}).

\bibitem[{\citenamefont{{Y. Nishida} and {H. Abuki}}(2005)}]{NA}
\bibinfo{author}{\bibnamefont{{Y. Nishida}}} \bibnamefont{and}
  \bibinfo{author}{\bibnamefont{{H. Abuki}}}, \bibinfo{journal}{Phys.\ Rev.\ D}
  \textbf{\bibinfo{volume}{72}}, \bibinfo{pages}{096004}
  (\bibinfo{year}{2005}).

\bibitem[{\citenamefont{{G. Sun} et~al.}(2007)\citenamefont{{G. Sun}, {L. He},
  and {P. Zhuang}}}]{SHZ}
\bibinfo{author}{\bibnamefont{{G. Sun}}}, \bibinfo{author}{\bibnamefont{{L.
  He}}}, \bibnamefont{and} \bibinfo{author}{\bibnamefont{{P. Zhuang}}},
  \bibinfo{journal}{Phys.\ Rev.\ D} \textbf{\bibinfo{volume}{75}},
  \bibinfo{pages}{096004} (\bibinfo{year}{2007}).

\bibitem[{\citenamefont{{M. Baldo} et~al.}(1995)\citenamefont{{M. Baldo}, {U.
  Lombardo}, and {P. Schuck}}}]{BLS}
\bibinfo{author}{\bibnamefont{{M. Baldo}}}, \bibinfo{author}{\bibnamefont{{U.
  Lombardo}}}, \bibnamefont{and} \bibinfo{author}{\bibnamefont{{P. Schuck}}},
  \bibinfo{journal}{Phys.\ Rev.\ C} \textbf{\bibinfo{volume}{52}},
  \bibinfo{pages}{975} (\bibinfo{year}{1995}).

\bibitem[{\citenamefont{{T. Tanigawa} and {M. Matsuzaki}}(1999)}]{Tani}
\bibinfo{author}{\bibnamefont{{T. Tanigawa}}} \bibnamefont{and}
  \bibinfo{author}{\bibnamefont{{M. Matsuzaki}}}, \bibinfo{journal}{Prog.\
  Theor.\ Phys.} \textbf{\bibinfo{volume}{102}}, \bibinfo{pages}{897}
  (\bibinfo{year}{1999}).

\bibitem[{\citenamefont{{M. Matsuo}}(2006)}]{Matsuo}
\bibinfo{author}{\bibnamefont{{M. Matsuo}}}, \bibinfo{journal}{Phys.\ Rev.\ C}
  \textbf{\bibinfo{volume}{73}}, \bibinfo{pages}{044309}
  (\bibinfo{year}{2006}).

\bibitem[{\citenamefont{{M. Matsuzaki}}(2000)}]{MM_q}
\bibinfo{author}{\bibnamefont{{M. Matsuzaki}}}, \bibinfo{journal}{Phys.\ Rev.\
  D} \textbf{\bibinfo{volume}{62}}, \bibinfo{pages}{017501}
  (\bibinfo{year}{2000}).

\bibitem[{\citenamefont{{H. Abuki} et~al.}(2002)\citenamefont{{H. Abuki}, {T.
  Hatsuda}, and {K. Itakura}}}]{AHI}
\bibinfo{author}{\bibnamefont{{H. Abuki}}}, \bibinfo{author}{\bibnamefont{{T.
  Hatsuda}}}, \bibnamefont{and} \bibinfo{author}{\bibnamefont{{K. Itakura}}},
  \bibinfo{journal}{Phys.\ Rev.\ D} \textbf{\bibinfo{volume}{65}},
  \bibinfo{pages}{074014} (\bibinfo{year}{2002}).

\bibitem[{\citenamefont{{M. Kitazawa} et~al.}(2004)\citenamefont{{M. Kitazawa},
  {T. Koide}, {T. Kunihiro}, and {Y. Nemoto}}}]{KKKN}
\bibinfo{author}{\bibnamefont{{M. Kitazawa}}},
  \bibinfo{author}{\bibnamefont{{T. Koide}}}, \bibinfo{author}{\bibnamefont{{T.
  Kunihiro}}}, \bibnamefont{and} \bibinfo{author}{\bibnamefont{{Y. Nemoto}}},
  \bibinfo{journal}{Phys.\ Rev.\ D} \textbf{\bibinfo{volume}{70}},
  \bibinfo{pages}{056003} (\bibinfo{year}{2004}).

\bibitem[{\citenamefont{{P. Castorina} et~al.}(2005)\citenamefont{{P.
  Castorina}, {G. Nardulli}, and {D. Zappal\`{a}}}}]{CNZ}
\bibinfo{author}{\bibnamefont{{P. Castorina}}},
  \bibinfo{author}{\bibnamefont{{G. Nardulli}}}, \bibnamefont{and}
  \bibinfo{author}{\bibnamefont{{D. Zappal\`{a}}}}, \bibinfo{journal}{Phys.\
  Rev.\ D} \textbf{\bibinfo{volume}{72}}, \bibinfo{pages}{076006}
  (\bibinfo{year}{2005}).

\bibitem[{\citenamefont{{T. Brauner}}(2008)}]{Br}
\bibinfo{author}{\bibnamefont{{T. Brauner}}}, \bibinfo{journal}{Phys.\ Rev.\ D}
  \textbf{\bibinfo{volume}{77}}, \bibinfo{pages}{096006}
  (\bibinfo{year}{2008}).

\bibitem[{\citenamefont{{Y. Nambu} and {G. Jona-Lasinio}}(1961)}]{NJL}
\bibinfo{author}{\bibnamefont{{Y. Nambu}}} \bibnamefont{and}
  \bibinfo{author}{\bibnamefont{{G. Jona-Lasinio}}}, \bibinfo{journal}{Phys.\
  Rev.} \textbf{\bibinfo{volume}{122}}, \bibinfo{pages}{345}
  (\bibinfo{year}{1961}).

\bibitem[{\citenamefont{{M. Gell-Mann} and {M. L{\`{e}}vy}}(1960)}]{GL}
\bibinfo{author}{\bibnamefont{{M. Gell-Mann}}} \bibnamefont{and}
  \bibinfo{author}{\bibnamefont{{M. L{\`{e}}vy}}}, \bibinfo{journal}{Il\ Nuovo\
  Cim.} \textbf{\bibinfo{volume}{16}}, \bibinfo{pages}{705}
  (\bibinfo{year}{1960}).

\bibitem[{\citenamefont{{L. He} et~al.}(2005)\citenamefont{{L. He}, {M. Jin},
  and {P. Zhuang}}}]{HJZ}
\bibinfo{author}{\bibnamefont{{L. He}}}, \bibinfo{author}{\bibnamefont{{M.
  Jin}}}, \bibnamefont{and} \bibinfo{author}{\bibnamefont{{P. Zhuang}}},
  \bibinfo{journal}{Phys.\ Rev.\ D} \textbf{\bibinfo{volume}{71}},
  \bibinfo{pages}{116001} (\bibinfo{year}{2005}).

\bibitem[{\citenamefont{{H. Mao} et~al.}(2006)\citenamefont{{H. Mao}, {N.
  Petropoulos}, {S. Shu}, and {W.-Q. Zhao}}}]{MPSZ}
\bibinfo{author}{\bibnamefont{{H. Mao}}}, \bibinfo{author}{\bibnamefont{{N.
  Petropoulos}}}, \bibinfo{author}{\bibnamefont{{S. Shu}}}, \bibnamefont{and}
  \bibinfo{author}{\bibnamefont{{W.-Q. Zhao}}}, \bibinfo{journal}{J.\ Phys.\ G}
  \textbf{\bibinfo{volume}{32}}, \bibinfo{pages}{2187} (\bibinfo{year}{2006}).

\bibitem[{\citenamefont{{J. O. Andersen}}(2007)}]{An}
\bibinfo{author}{\bibnamefont{{J. O. Andersen}}}, \bibinfo{journal}{Phys.\
  Rev.\ D} \textbf{\bibinfo{volume}{75}}, \bibinfo{pages}{065011}
  (\bibinfo{year}{2007}).

\bibitem[{\citenamefont{{J. Deng} et~al.}(2007)\citenamefont{{J. Deng}, {A.
  Schmitt}, and {Q. Wang}}}]{DSW}
\bibinfo{author}{\bibnamefont{{J. Deng}}}, \bibinfo{author}{\bibnamefont{{A.
  Schmitt}}}, \bibnamefont{and} \bibinfo{author}{\bibnamefont{{Q. Wang}}},
  \bibinfo{journal}{Phys.\ Rev.\ D} \textbf{\bibinfo{volume}{76}},
  \bibinfo{pages}{034013} (\bibinfo{year}{2007}).

\bibitem[{\citenamefont{{W. Broniowski} and {W. Florkowski}}(2001)}]{BF}
\bibinfo{author}{\bibnamefont{{W. Broniowski}}} \bibnamefont{and}
  \bibinfo{author}{\bibnamefont{{W. Florkowski}}}, \bibinfo{journal}{Phys.\
  Rev.\ Lett.} \textbf{\bibinfo{volume}{87}}, \bibinfo{pages}{272302}
  (\bibinfo{year}{2001}).

\bibitem[{\citenamefont{{S. Hands} et~al.}(2005)\citenamefont{{S. Hands}, {S.
  Kim}, and {J.-I. Skullerud}}}]{HKS}
\bibinfo{author}{\bibnamefont{{S. Hands}}}, \bibinfo{author}{\bibnamefont{{S.
  Kim}}}, \bibnamefont{and} \bibinfo{author}{\bibnamefont{{J.-I. Skullerud}}},
  \bibinfo{journal}{PoS LAT2005} p. \bibinfo{pages}{149}
  (\bibinfo{year}{2005}).

\bibitem[{\citenamefont{{J. B. Kogut} and {D. K.
  Sinclair}}(2002{\natexlab{b}})}]{KS2}
\bibinfo{author}{\bibnamefont{{J. B. Kogut}}} \bibnamefont{and}
  \bibinfo{author}{\bibnamefont{{D. K. Sinclair}}}, \bibinfo{journal}{Phys.\
  Rev.\ D} \textbf{\bibinfo{volume}{66}}, \bibinfo{pages}{014508}
  (\bibinfo{year}{2002}{\natexlab{b}}).

\bibitem[{\citenamefont{{J. I. Kapusta}}(1989)}]{Ka}
\bibinfo{author}{\bibnamefont{{J. I. Kapusta}}},
  \emph{\bibinfo{title}{Finite-Temperature Field Theory}}
  (\bibinfo{publisher}{Cambridge University Press}, \bibinfo{year}{1989}).

\bibitem[{\citenamefont{{D. Ebert} and {K. G. Klimenko}}(2006)}]{EK}
\bibinfo{author}{\bibnamefont{{D. Ebert}}} \bibnamefont{and}
  \bibinfo{author}{\bibnamefont{{K. G. Klimenko}}}, \bibinfo{journal}{Eur.\
  Phys.\ J.\ C} \textbf{\bibinfo{volume}{46}}, \bibinfo{pages}{771}
  (\bibinfo{year}{2006}).

\bibitem[{\citenamefont{{H. Abuki} et~al.}(2009)}]{Abu}
\bibinfo{author}{\bibnamefont{{H. Abuki}}} \bibnamefont{et~al.},
  \bibinfo{journal}{Phys.\ Rev.\ D} \textbf{\bibinfo{volume}{79}},
  \bibinfo{pages}{034032} (\bibinfo{year}{2009}).

\bibitem[{\citenamefont{{X. Hao} and {P. Zhuang}}(2007)}]{HZ}
\bibinfo{author}{\bibnamefont{{X. Hao}}} \bibnamefont{and}
  \bibinfo{author}{\bibnamefont{{P. Zhuang}}}, \bibinfo{journal}{Phys.\ Lett.\
  B} \textbf{\bibinfo{volume}{652}}, \bibinfo{pages}{275}
  (\bibinfo{year}{2007}).

\bibitem[{\citenamefont{{L. P. Gor'kov}}(1958)}]{Go}
\bibinfo{author}{\bibnamefont{{L. P. Gor'kov}}}, \bibinfo{journal}{Sov.\ Phys.\
  JETP} \textbf{\bibinfo{volume}{34}}, \bibinfo{pages}{735}
  (\bibinfo{year}{1958}).

\bibitem[{\citenamefont{{F. Pistolesi} and {G. C. Strinati}}(1994)}]{PS}
\bibinfo{author}{\bibnamefont{{F. Pistolesi}}} \bibnamefont{and}
  \bibinfo{author}{\bibnamefont{{G. C. Strinati}}}, \bibinfo{journal}{Phys.\
  Rev.\ B} \textbf{\bibinfo{volume}{49}}, \bibinfo{pages}{6356}
  (\bibinfo{year}{1994}).

\bibitem[{\citenamefont{{M. Matsuzaki}}(1998)}]{MM_D}
\bibinfo{author}{\bibnamefont{{M. Matsuzaki}}}, \bibinfo{journal}{Phys.\ Rev.\
  C} \textbf{\bibinfo{volume}{58}}, \bibinfo{pages}{3407}
  (\bibinfo{year}{1998}).

\bibitem[{\citenamefont{{Y. Nambu}}(1960)}]{Nam}
\bibinfo{author}{\bibnamefont{{Y. Nambu}}}, \bibinfo{journal}{Phys.\ Rev.}
  \textbf{\bibinfo{volume}{117}}, \bibinfo{pages}{648} (\bibinfo{year}{1960}).

\bibitem[{\citenamefont{{P. Camiz} et~al.}(1966)\citenamefont{{P. Camiz}, {A.
  Covello}, and {M. Jean}}}]{CCJ}
\bibinfo{author}{\bibnamefont{{P. Camiz}}}, \bibinfo{author}{\bibnamefont{{A.
  Covello}}}, \bibnamefont{and} \bibinfo{author}{\bibnamefont{{M. Jean}}},
  \bibinfo{journal}{Il\ Nuovo\ Cim.\ B} \textbf{\bibinfo{volume}{42}},
  \bibinfo{pages}{199} (\bibinfo{year}{1966}).

\bibitem[{\citenamefont{{M. Loewe} and {C. Villavicencio}}(2004)}]{LV2}
\bibinfo{author}{\bibnamefont{{M. Loewe}}} \bibnamefont{and}
  \bibinfo{author}{\bibnamefont{{C. Villavicencio}}}, \bibinfo{journal}{Phys.\
  Rev.\ D} \textbf{\bibinfo{volume}{70}}, \bibinfo{pages}{074005}
  (\bibinfo{year}{2004}).

\bibitem[{\citenamefont{{G. Colangelo} et~al.}(2001)\citenamefont{{G.
  Colangelo}, {J. Gasser}, and {H. Leutwyler}}}]{CGL}
\bibinfo{author}{\bibnamefont{{G. Colangelo}}},
  \bibinfo{author}{\bibnamefont{{J. Gasser}}}, \bibnamefont{and}
  \bibinfo{author}{\bibnamefont{{H. Leutwyler}}}, \bibinfo{journal}{Nucl.\
  Phys.\ B} \textbf{\bibinfo{volume}{603}}, \bibinfo{pages}{125}
  (\bibinfo{year}{2001}).

\bibitem[{\citenamefont{{B. B. Back} et~al.}(2000)}]{Ba}
\bibinfo{author}{\bibnamefont{{B. B. Back}}} \bibnamefont{et~al.},
  \bibinfo{journal}{Phys.\ Rev.\ Lett.} \textbf{\bibinfo{volume}{85}},
  \bibinfo{pages}{3100} (\bibinfo{year}{2000}).

\bibitem[{\citenamefont{{A. Ayala} and {A. Smerzi}}(1997)}]{AS}
\bibinfo{author}{\bibnamefont{{A. Ayala}}} \bibnamefont{and}
  \bibinfo{author}{\bibnamefont{{A. Smerzi}}}, \bibinfo{journal}{Phys.\ Lett.\
  B} \textbf{\bibinfo{volume}{405}}, \bibinfo{pages}{20}
  (\bibinfo{year}{1997}).

\bibitem[{\citenamefont{{S.-T. Wu} and {S. Yip}}(2003)}]{WY}
\bibinfo{author}{\bibnamefont{{S.-T. Wu}}} \bibnamefont{and}
  \bibinfo{author}{\bibnamefont{{S. Yip}}}, \bibinfo{journal}{Phys.\ Rev.\ A}
  \textbf{\bibinfo{volume}{67}}, \bibinfo{pages}{053603}
  (\bibinfo{year}{2003}).

\end{thebibliography}
\end{document}